\documentclass[prl,aps,showpacs,showkeys,groupedaddress,superscriptaddress,twocolumn]{revtex4-1}

\usepackage[utf8x]{inputenc}
\usepackage{color}
\usepackage{bbm} 
\usepackage{amsfonts,amsmath,amssymb,stmaryrd}
\usepackage{graphicx}
\usepackage{subfigure}  % use for side-by-side figures
\usepackage{hyperref}
\usepackage{epsfig}
\usepackage{mathrsfs}
\usepackage{verbatim}
\usepackage{centernot}
\usepackage{float}

\newcommand{\bra}[1]{\langle#1|}
\newcommand{\ket}[1]{|#1\rangle}

\renewcommand{\c}{\hat{c}}

\usepackage{array}

\usepackage{bm}	% bold in math mode

\begin{document}
\title{Reservoir-induced Thouless pumping and symmetry protected topological order in open quantum chains}

\author{D. Linzner}
\affiliation{Department of Physics and Research Center OPTIMAS, University of Kaiserslautern, Germany}
\author{L. Wawer}
\affiliation{Department of Physics and Research Center OPTIMAS, University of Kaiserslautern, Germany}
\author{F. Grusdt}
\affiliation{Department of Physics, Harvard University, Cambridge, Massachusetts 02138, USA}
\author{M. Fleischhauer}
\affiliation{Department of Physics and Research Center OPTIMAS, University of Kaiserslautern, Germany}

\begin{abstract}
We introduce a classification scheme for symmetry protected topological phases applicable to stationary states of open systems based on a generalization of the many-body polarization. The polarization can be used to probe the topological properties of non-interacting and interacting closed and open systems as well
and remains a meaningful quantity even in the presence of moderate particle-number fluctuations.
As examples, we discuss two open-system versions of a topological Thouless pump
in the steady state of one-dimensional lattices driven by Markovian reservoirs.
In the analogous unitary system, the Rice-Mele model, symmetries enforce topological properties which lead to a non-trivial winding of the geometric Zak phase upon cyclic variations of model parameters. Associated with this is a winding of the many-body polarization, corresponding to a quantized transport in the bulk (Thouless pump). We here show that in the open system, where the Zak phase looses its meaning, the same symmetries enforce a winding of the generalized many-body polarization. This winding is shown to be 
robust against Hamiltonian perturbations as well as homogeneous dephasing and particle losses.
\end{abstract}

\date{\today}
\maketitle

%%%%%%%%%%%%%%%%%%%%%%%%%%%%%%%%%%%%
\paragraph{Introduction.—} 
%%%%%%%%%%%%%%%%%%%%%%%%%%%%%%%%%%%
%
Since the discovery of the quantum Hall effect \cite{Q-Hall}  topological states of matter have fascinated scientists in all fields of physics.
Topological phases can be characterized by non-local integer invariants \cite{Thouless-PRL-1982}, whose existence leads to a number of fascinating 
properties including protected edge states \cite{Hatsugai-1993} and anyonic excitations \cite{Arovas}.
Their robustness against disorder has made topological systems important tools in metrology and promising candidates for future quantum information platforms \cite{top-QC}. However, topological protection is typically destroyed by dissipation, e.g. decoherence or particle losses \cite{Budich-2012}. 
It would thus be highly desirable to extend the notion of topological order to open systems coupled to external reservoirs.
This could even substantially extend topological protection since steady states are attractors of the dynamics and thus have an intrinsic robustness. 

By now there is a rather good understanding of topological order in non-interacting closed systems. An exhaustive classification 
of gapped topological states of non-interacting fermions 
(the "ten-fold way" \cite{Schnyder-PRB-2008}) has been given solely on the basis of general 
symmetry properties.
In contrast, the understanding of topological order in interacting systems \cite{Wen-1995} is limited and the extension to open systems is entirely in its infancy. 
Recently there have been some 
attempts to generalize the concept of topological order to 
steady states  of open systems 
\cite{Diehl-Nat-Phys-2008,Diehl-Nat-Phys-2011,Albert-arxiv-2015}. This includes proposals to extend Berry phases 
\cite{Berry1984,Zak-PRL-1989,Xiao-RMP-2010} to mixed states \cite{Avron-NJP-2011,Rivas-PRB-2013,Viyuela-PRL-2014,Huang-PRL-2014}
e.g. using the 
Uhlmann phase \cite{Uhlmann-Rep-Math-Phys-1986}, whose suitability has however been questioned
\cite{Budich-arxiv-2015}. Alternatives suggested for one-dimensional systems are based on
projective symmetry group representations \cite{Nieuwenburg-PRB-2014}.
For gaussian density matrices describing steady states of non-interacting fermionic systems with linear reservoir couplings, 
 a classification of topological order can be made \cite{Bardyn-NJP-2013} employing a scheme similar to the unitary case \cite{Schnyder-PRB-2008}.  

 %%%%%%%%%%%%%%%%%%%%%%%%%%%%%%%%%%%%%%%%%%%%%%%%%%%%%
\begin{figure}[htb]
\centering
\epsfig{file=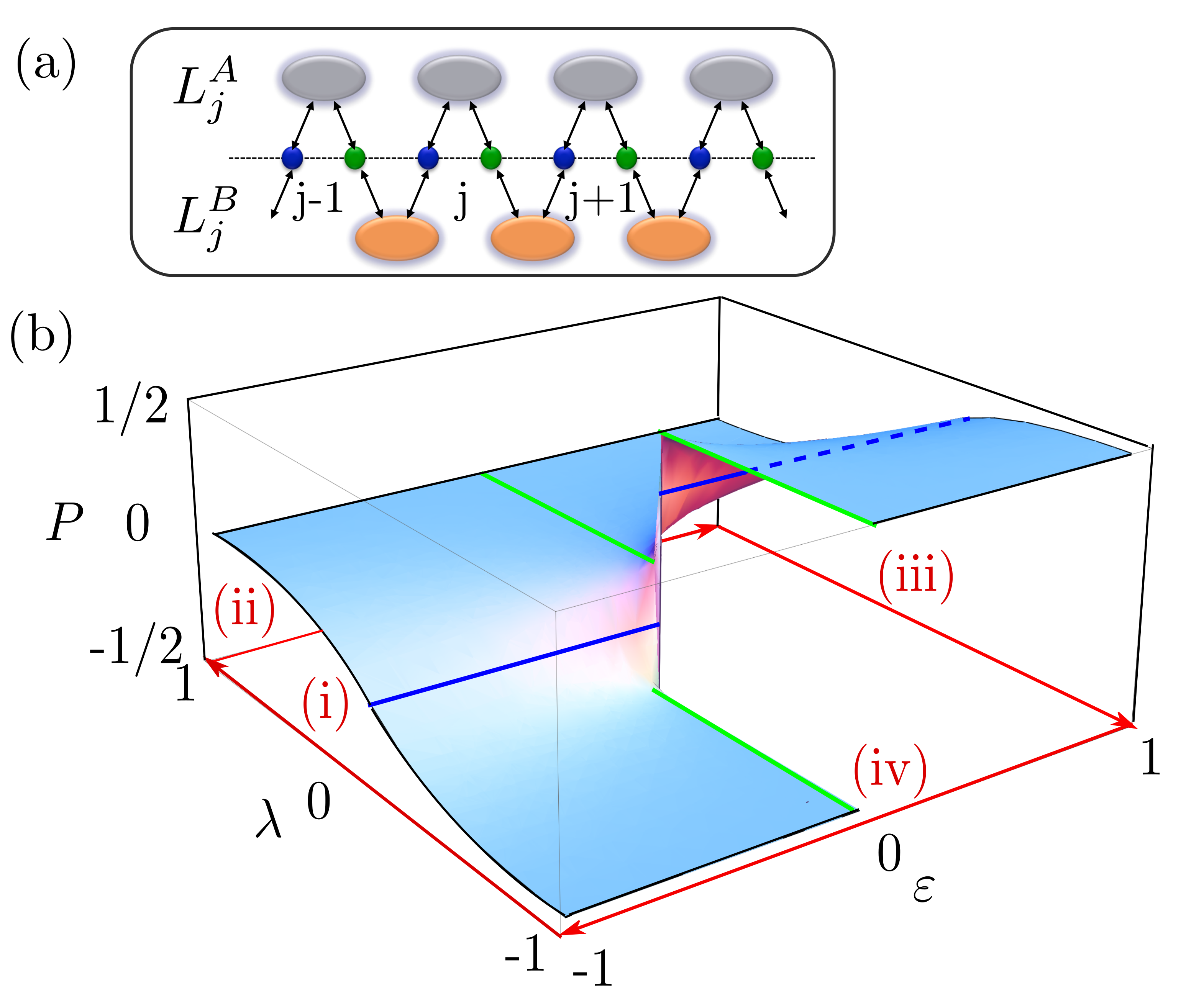, width=0.48\textwidth}
\caption{(Color online) (a) Open-system analogue of the Rice-Mele model: One-dimensional fermion or spin chain with alternating pairwise coupling to two different Markovian reservoirs (Lindblad generators $L_{j}^A$ and $L_{j}^B$). The unit cell labeled by $"j"$ contains a left "$L$"(blue) and a right "$R$"(green) lattice site. (b) Winding of the many-body polarization in the fermionic model (see text) upon cyclic variations of parameters of the Liouvillian. On the axes with inversion symmetry (blue and green) the change in polarization is strictly quantized $\Delta P=0, \pm 1/2$. On the special path along the periphery  (red), the steady state is pure.
 }
\label{fig:system}
\end{figure}
%%%%%%%%%%%%%%%%%%%%%%%%%%%%%%%%%%%%%%%%%%%%%%%%%%%%%

Here we pursue a different and more general approach, applicable also to interacting open systems.
We consider the case where topology is
protected by certain symmetries (symmetry protected topological order, SPT). 
A prime example of a unitary system with SPT 
is the Rice-Mele model \cite{Rice-Mele-PRL-1982}. The symmetries of the model enforce a non-trivial winding of the geometric Berry (or Zak) phase \cite{Berry1984,Zak-PRL-1989} upon cyclic variations of the parameters, which defines a topological invariant, the Chern number. As shown 
in \cite{King-Smith,Ortiz-1994,Xiao-RMP-2010} the winding of the Zak phase is equivalent to a winding of the polarization, giving rise to a quantized particle transport known as Thouless pump
\cite{Thouless-PRL-1982,Thouless-PRB-1983, Lohse-2015, Nakajima-2016}. Such topological pumps are of special interest as they allow for studying the topological structure of parameter spaces \cite{Berg-2011} and have been used to understand $Z_2$ topological insulators \cite{Fu-PRB-2006}. Here we 
show that topological pumps can be used to classify SPT in the steady state of open, one-dimensional models. 
Specifically we consider open chains of fermions and spins with
a Liouvillian dynamics obeying the same symmetries as the Rice-Mele model. We show that they enforce a non-trivial
winding of the many-body polarization \cite{Resta-PRL-1998}, characterizing a Thouless pump.
While the Zak phase is no longer suitable to characterize topology, the winding of the polarization generalized to open systems with moderate particle number fluctuations defines a quantized topological invariant.

%%%%%%%%%%%%%%%%%%%%%%%%%%%%%%%%%%%
\paragraph{Classification scheme of SPT.—}
%%%%%%%%%%%%%%%%%%%%%%%%%%%%%%%%%%%
In the following we want to introduce our classification scheme, which generalizes the notion of SPT to open systems. It is motivated by the equivalence between the Zak phase and the 
many-body polarization \cite{King-Smith} formulated for periodic systems
by Resta \cite{Resta-PRL-1998} as
\begin{equation}
P = \frac{1}{2\pi} \operatorname{Im}\, \ln \biggl\langle \exp\biggl\{ i\frac{2\pi}{L}\hat{X}\biggr\}\biggr\rangle,
\label{eq:Resta}
\end{equation}
with $\hat{X}$ being the center of mass operator.
This quantity has already been successfully applied to  characterize quantized topological transport and SPT in one dimensional $L$-periodic closed systems  \cite{Xiao-RMP-2010}. We here argue that this concept can be further generalized to periodic systems subject to dissipation. 

In fact for this quantity to be meaningful the only restriction to be made is to fix the mean particle-number $\langle \hat{N} \rangle= L$  with non-extensive fluctuations, i.e. $\langle\Delta \hat{N}\rangle/\langle \hat{N} \rangle \to 0$ for all relevant states when $L \to \infty$. Under these conditions 
$P$, which is defined mod $1$, is invariant under trivial changes of the particle coordinates by one unit cell in the limit $L \to \infty$.
For half filling the system has furthermore particle-hole symmetry (PHS). We note that $P$ contains genuine many-particle contributions and is in general not determined by single-particle correlations alone. 

We can now use eq.\eqref{eq:Resta} to study systems with additional symmetries and a unique stationary state. Focusing concretely on spatial inversion (IS), we consider two inversion symmetric systems with polarizations $P_1$ and $P_2$. Because under inversion the difference of their polarizations $\Delta P=P_1-P_2$ changes its sign $\Delta P\rightarrow -\Delta P$, but at the same time has to remain invariant because of IS, it follows that 
%\begin{align}
$\Delta P =-\Delta P \ \text{mod} \ 1.$
%\end{align}
Therefore the difference $\Delta P$ can only take two quantized values $\Delta P = 0, 1/2$ \cite{Zak-PRL-1989}. This quantization of the polarization has been heavily used in the context of closed periodic systems, e.g. in the Su-Schrieffer-Heeger (SSH) model \cite{SSH}, where the polarization can be directly related to the Zak phase \cite{King-Smith} and underlies the "ten-fold way" \cite{Schnyder-PRB-2008}. We stress that this already makes it possible to classify SPT-phases in non-interacting and interacting closed and open systems as well.

In the following we demonstrate our method with two toy-models. We further prove that these systems can show a non-trivial winding number of the polarization on closed loops in parameter space, extending the concept of topological pumps to open systems. 

%%%%%%%%%%%%%%%%%%%%%%%%%%%%%%%%%%%
\paragraph{Model.—}
%%%%%%%%%%%%%%%%%%%%%%%%%%%%%%%%%%%
We investigate driven-dissipative analogues of the translationally invariant Rice-Mele (RM) model \cite{Rice-Mele-PRL-1982}.
As indicated in Fig.\ref{fig:system}a) we consider one-dimensional lattices of fermions and spins coupled to Markovian reservoirs described by Lindblad operators
\begin{eqnarray}
L_j^A &=& \sqrt{\Gamma(1+\varepsilon)}\biggl[ (1-\lambda)\left( \hat A_{L,j}^\dagger +\hat A_{R,j}\right) +\nonumber\\
&& \qquad\qquad\quad+ (1+\lambda)\left(\hat A_{L,j}\mp \hat A_{R,j}^\dagger\right)\biggr], \label{eq:Lindblad_A}\\
L_j^B &=& \sqrt{\Gamma(1-\varepsilon)}\biggl[ (1-\lambda)\left( \hat A_{L,j+1}^\dagger +\hat A_{R,j}\right) +\nonumber\\
&& \qquad\qquad\quad + (1+\lambda)\left(\hat A_{L,j+1}\mp \hat A_{R,j}^\dagger\right)\biggr].\label{eq:Lindblad_B}
\end{eqnarray}
Each unit cell is defined and numbered by the index "$j$" and consists of two sites "\textit{L}" and "\textit{R}". 
$\hat A$ and $\hat A^\dagger$ denote fermion annihilation and creation operators, $\c,\c^\dagger$, (minus sign), or spin-operators
$\hat\sigma^-,\hat\sigma^+$ (plus sign) respectively.
$\Gamma$ is the coupling strength to the reservoirs and determines the overall time scale. 
Our model is characterized by two parameters $\lambda\in[-1,1]$ and $\varepsilon\in[-1,1]$, where $\varepsilon$ describes the relative strength of reservoir couplings % acting 
across inequivalent links in the lattice and $\lambda$ controls the distribution of particles inside a unit cell.
% in the steady state. 
For $\lambda =1$ ($\lambda = -1$) both Lindblad operators localize particles on the $L$-sites ($R$-sites). 
% Vice versa for $\lambda = -1$ the reservoirs drive spin excitations to the $R$-sites. 
Despite their model character, there have been proposals for experimental realizations of reservoirs of similar structure using cold atoms \cite{Diehl-Nat-Phys-2008,Bardyn-NJP-2013,Hoening-2011}. 

We study the steady state  $\rho_{\rm ss}$ of the systems density matrix, which obeys the Master equation in Lindblad form ($\hbar =1$)
$\partial_t \rho_{\rm ss} = {\cal L}\rho_{\rm ss}=0$, with
\begin{equation}
{\cal L}\rho  = -i \bigl[H,\rho\bigr]\! +\!\frac{1}{2} \sum_\mu \left(2 L_\mu \rho L_\mu^\dagger -L_\mu^\dagger L_\mu \rho -\rho L_\mu^\dagger L_\mu \right),
\end{equation}
where $L_\mu\in \left(L_j^A, L_j^B\right)$. We consider periodic boundary conditions with $j=1,\dots, L$.
To ensure a unique steady state for all parameters we introduce a potential,
$H= U\sum_j\left(\hat A_{L,j}^{\dagger}\hat A_{L,j}\pm \hat A_{R,j}^{\dagger}\hat A_{R,j}\right)$ for fermions ($+$) or spins ($-$) respectively and set $U=\Gamma$ for convenience.
We further have to add non-linear modifications in the case of spins
\begin{align}
	L^A_j&\rightarrow L^A_j+\sqrt{\Gamma(1+\varepsilon)}\left(\hat{\sigma}_{L,j}^+\hat{\sigma}^+_{R,j}-\hat{\sigma}^-_{L,j}\hat{\sigma}^-_{R,j}\right)\label{eq:nonlin_A},\\
	L^B_j&\rightarrow  L^B_j\!+\!\sqrt{\Gamma(1-\varepsilon)}\left(\hat{\sigma}_{R,j}^+\hat{\sigma}_{L,j+1}^+-\hat{\sigma}^-_{R,j}\hat{\sigma}^-_{L,j+1}\right)\label{eq:nonlin_B},
\end{align} 
in order to prevent the steady state from becoming completely mixed on the $\lambda =0$ axis (see supplementary material for details).
We note that while the fermionic chain is gaussian, the spin chain is an example for a non-gaussian, i.e. an interacting system.

The following analysis relies on symmetries of the Liouvillians. Besides translation invariance by one unit cell, ${\cal L}$
is invariant under $C\cdot I_b$, i.e. simultaneous particle-hole exchange $(C)$ and spatial inversion with respect to any lattice bond $(I_b)$. 
%, as e.g. the transformation  $\hat \sigma_{L,j-n}^- \leftrightarrow \hat\sigma_{R,j+n}^+$ for $n=0,1,2,...$ leaves ${\cal L}$ invariant. 
As a consequence one finds that the
average number of particles per unit cell  is unity, $\langle\hat n_{L,j} +\hat n_{R,j}\rangle=1$, corresponding to half filling or PHS. 
For the discussion of symmetry protected topological order we furthermore note that the
system is invariant with respect to an inversion at a lattice site $(I_s)$ for $\varepsilon=0$ and with respect to a bond $(I_b)$ for $\lambda =0$.

%%%%%%%%%%%%%%%%%%%%%%%%%%%%%%%%%%%
\paragraph{SPT and winding of polarization.—}
%%%%%%%%%%%%%%%%%%%%%%%%%%%%%%%%%%%

We now show that the symmetries of the system enforce 
a non-trivial winding of the polarization upon a closed loop in the two dimensional parameter space $(\varepsilon,\lambda)$. 
To see this we note that 
 $\lambda \to -\lambda$ is equivalent to the action of $C$,
and $\varepsilon \to -\varepsilon$ to $I_s$. The most important consequence is that simultaneous transformation $(\lambda ,\varepsilon)\to (-\lambda,-\varepsilon)$ 
is equivalent to a translation by one lattice site, i.e. $(L,j)\to (R,j)$ and $(R,j)\to(L,j+1)$.
Thus performing a point-inversion in parameter space, i.e.  $(\lambda,\varepsilon) \to (-\lambda,-\varepsilon)$ leaves ${\cal L}$ invariant up to a translation by one lattice site and if the polarization is well defined we find
\begin{align}
P(-\lambda,-\varepsilon)=P(\lambda,\varepsilon)\pm 1/2.\label{eq:mirror}
\end{align}
As the steady state is unique, the change in polarization after one cycle has to be integer quantized, which, together with \eqref{eq:mirror} fixes the winding number of the polarization to $\nu=\pm 1,\pm 3,... $
This is confirmed by semi-analytical and numerical TEBD calculations for both types of chains, see Fig. \ref{fig:system} b).

Furthermore \eqref{eq:mirror} also results in the existence of two different SPT phases with a difference in polarization $\Delta P=1/2$ respectively for general parameter values on the $\lambda=0$- or $\varepsilon=0$-axis, see Fig.\ref{fig:system} b). Therefore a topological transition takes place in the center $\lambda=\varepsilon=0$, where $P$ is not defined and $\Delta P$ can change discontinuously.

%%%%%%%%%%%%%%%%%%%%%%%%%%%%%%%%%
\paragraph{Pure steady states.—}
%%%%%%%%%%%%%%%%%%%%%%%%%%%%%%%%%
%
It is interesting to consider the steady state 
% of the Liouvillian 
for parameters along
the special path in parameter space indicated in Fig.\ref{fig:system} b):
(i) $\varepsilon= 1$, $\lambda = 1 \to -1$,  (ii) $\lambda = -1$, $\varepsilon = 1 \to -1$, (iii)
$\varepsilon=-1$, $\lambda = -1 \to  +1$, and finally (iv) $\lambda = +1$, $\varepsilon=-1 \to 1$.
% For the parameter path (i)  - (iv) 
Here a pure steady state $\rho_{ss}=|\Psi\rangle\langle\Psi|$ exists.
On  (i) and (ii)  it factorizes in unit cells $|\Psi(\lambda)\rangle = \prod_j |\phi(\lambda)\rangle_j$, where up to normalization
 $ |\phi(\lambda)\rangle_j = (1-\lambda) \hat{A}^\dagger_{L,j}|0\rangle -(1+\lambda) \hat{A}^\dagger_{R,j} |0\rangle$.
Here $|0\rangle$ denotes $|\downarrow_{L,j}\downarrow_{R,j}\rangle$ for spins and the vacuum for fermions.
On parts (iii) and (iv) the steady state factorizes in new unit cells that are shifted by one site
with respect to the old ones.

Because $|\Psi\rangle$ is a pure steady state (dark state), it is the ground state of the parent Hamiltonian
$H_{\rm par} = \sum_\mu  L_\mu^\dagger L_\mu \ge 0$  which encodes all properties of the steady state. 
For our models with periodic boundary conditions the parent Hamiltonian reads
$H _{\rm par}= H_{\rm RM} + H_1$, where
$ H_{\rm RM}$
is the Rice-Mele Hamiltonian, describing the dynamics of fermions or spin excitations in a superlattice with alternating hopping matrix elements 
$t_1 = 2 \Gamma (1+\varepsilon)(1-\lambda^2)$ and $t_2 = 2 \Gamma(1-\varepsilon) (1-\lambda^2)$ and a staggered potential $\pm \Delta$ with $\Delta = 8 \lambda \Gamma $.
This model is a paradigmatic example for a topological Thouless pump \cite{Thouless-PRB-1983}. When following the systems ground state adiabatically while $t_1-t_2$ and $\Delta$ encircle the topological singularity at $t_1=t_2$, $\Delta=0$, a quantized current is induced in the bulk. It can be associated with a quantized winding of the many-body polarization, corresponding to an integer topological Chern number $(C=1)$. The second term $H_1$ has no effect on the dark state since $H_1 |\Psi(\lambda)\rangle =0$.

%%%%%%%%%%%%%%%%%%%%%%%%%%%%%%%%%%%
\paragraph{Robustness.—}
%%%%%%%%%%%%%%%%%%%%%%%%%%%%%%%%%%%
%
%%%%%%%%%%%%%%%%%%%%%%%%%%%%%%%%%%%%%%%%%%%%%%%%%%%%%
\begin{figure}[t]
\centering
\epsfig{file=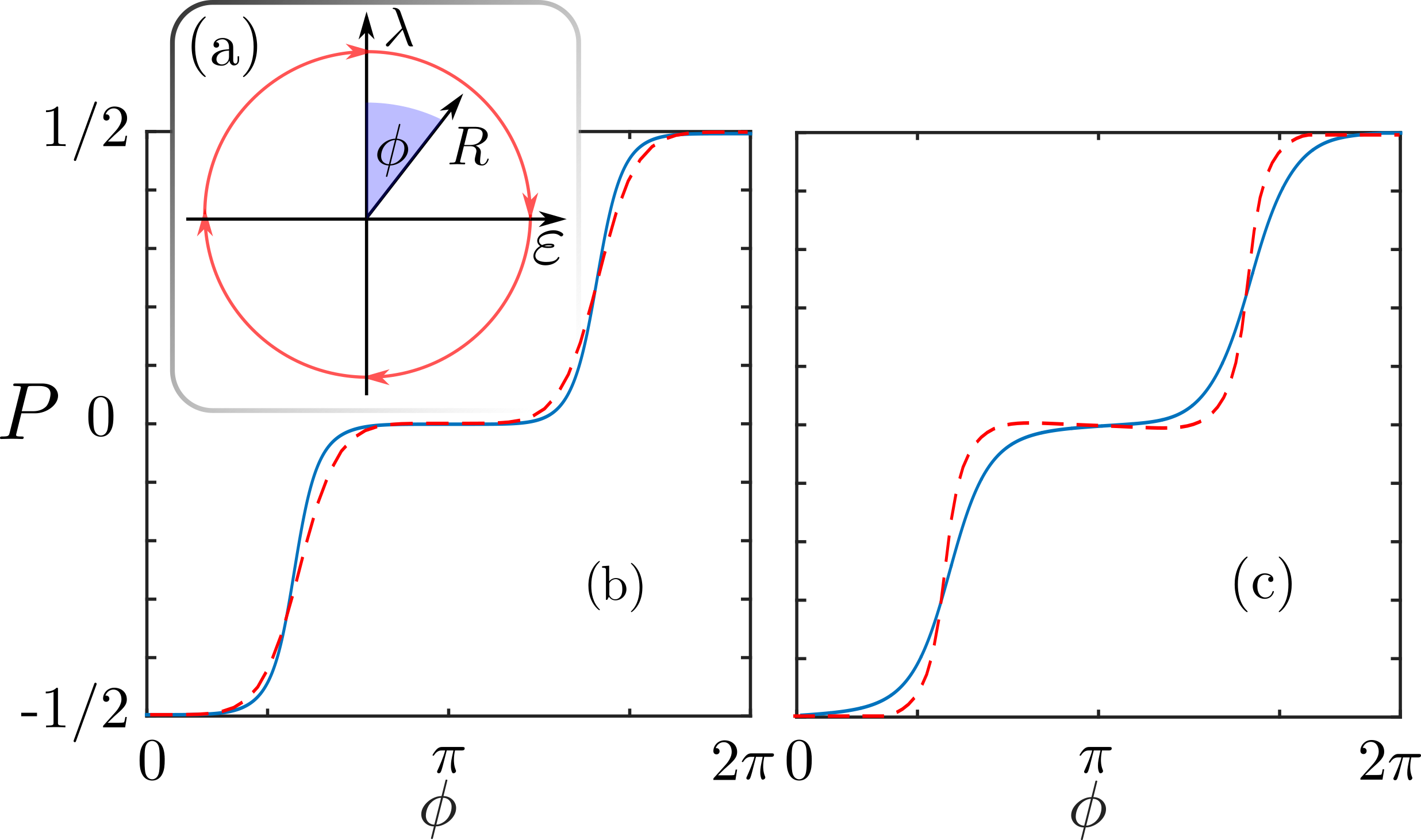, width=0.42\textwidth}
\caption{(Color online) Many-body polarization under slow variation of the parameters evaluated along 
a circular path (red) as indicated in (a) with radius $R=\sqrt{\varepsilon^2+\lambda^2}=1$. Different perturbations are considered numerically, using TEBD for the spin-chain ($L=16$ unit cells) or exact diagonalization for the four-site periodic fermionic-chain ($L=2$ unit cells)(dashed):
 (b) spatially random magnetic field in $z$ direction $V/\Gamma =1$ (disorder average consisting of 50 samples).
 (c) local dephasing with homogeneous losses and
 pumping both with rates $\gamma_\perp/\Gamma=0.5$, where $\gamma_\parallel=\gamma_\perp$.}
\label{fig:robustness}
\end{figure}
%%%%%%%%%%%%%%%%%%%%%%%%%%%%%%%%%%%%%%%%%%%%%%%%%%%%%
%
In closed systems topological properties are known to be robust against perturbations that do not close the energy gap.
% and conserve the number of particles.
This makes it interesting to ask whether the winding of the polarization in an open system can be robust against Hamiltonian and Liouvillian perturbations. 

For the spin chain we evolved the stationary state numerically, employing the time-dependent block decimation (TEBD)\cite{Vidal2004} to investigate robustness in sufficiently large systems ($L=16$) along a circular path as indicated in Fig.\ref{fig:robustness} a). 

For the fermionic chain, which is linear in fermionic operators, we can calculate the polarization of its steady state semi-analytically \cite{Eisert-Proszen}. However because we add perturbations that are non-linear in fermionic operators like dephasing or spatial randomness, we can only check for robustness in small periodic systems of four sites. This is sufficient because topological effects should not depend on system size.

Specifically we analyzed three different types of perturbations which preserve the PHS of the steady state: (i) a spatially random
magnetic field in the $z$-direction with uniform distribution, (ii) dephasing, and (iii) additional local losses compensated by local pumps with equal rates.
In the first case we add a Hamiltonian 
 $H_{\rm pert} = \sum_\mu V w_\mu \left(1-2\hat{A}_\mu^\dagger\hat{A}_\mu\right)$, where $\mu$ denotes a lattice site and $w_\mu\in[-1,1]$ are 
independent random magnetic field strengths.  As can be seen from Fig.\ref{fig:robustness} b) this perturbation has negligible influence on the polarization and does not affect its winding.
In the second case we add an additional Lindblad term with
$L^{R/L,\mu}_{\rm deph} = \sqrt{\gamma_\perp} \, \left(1-2\hat{A}_{L/R,\mu}^\dagger\hat{A}_{L/R,\mu}\right)$, where $\gamma_\perp$ is the dephasing rate.
This term, too, has almost no effect on $P$ and does not change its winding. The same is true  for homogeneous losses and gain. These have to be added with equal rates $\gamma_\parallel$ however in order to maintain the PHS of the system, i.e. $\langle \hat{n}_L+\hat{n}_R\rangle=1$: $L^{R/L,\mu}_\uparrow = \sqrt{\gamma_\parallel} \, \hat{A}_{R/L,\mu}^\dagger$, and 
$L^{R/L,\mu}_\downarrow = \sqrt{\gamma_\parallel} \, \hat{A}_{R/L,\mu}$. The small effects of these non-hermitian perturbations are shown in Fig.\ref{fig:robustness} c).

Most interestingly $\Delta P =\pm 1/2$ between two inversion-symmetric points on the $\lambda=0$ axis
remains intact even in the presence of spatially homogeneous losses (see Fig.\ref{fig:robustness} c)). These findings are important, because they show that in an open system topological properties such as the quantized charge pump
can be maintained even in the presence of unwanted losses, if the system remains PHS.
% at least as long as these losses are compensated by an appropriate gain term. 

%%%%%%%%%%%%%%%%%%%%%%%%%%%%%%%%%%
\paragraph{Topological singularity.—}
%%%%%%%%%%%%%%%%%%%%%%%%%%%%%%%%%%%
%
%%%%%%%%%%%%%%%%%%%%%%%%%%%%%%%%%%%%%%%%%%%%%%%%%%%%
\begin{figure}[t]
\centering
\epsfig{file=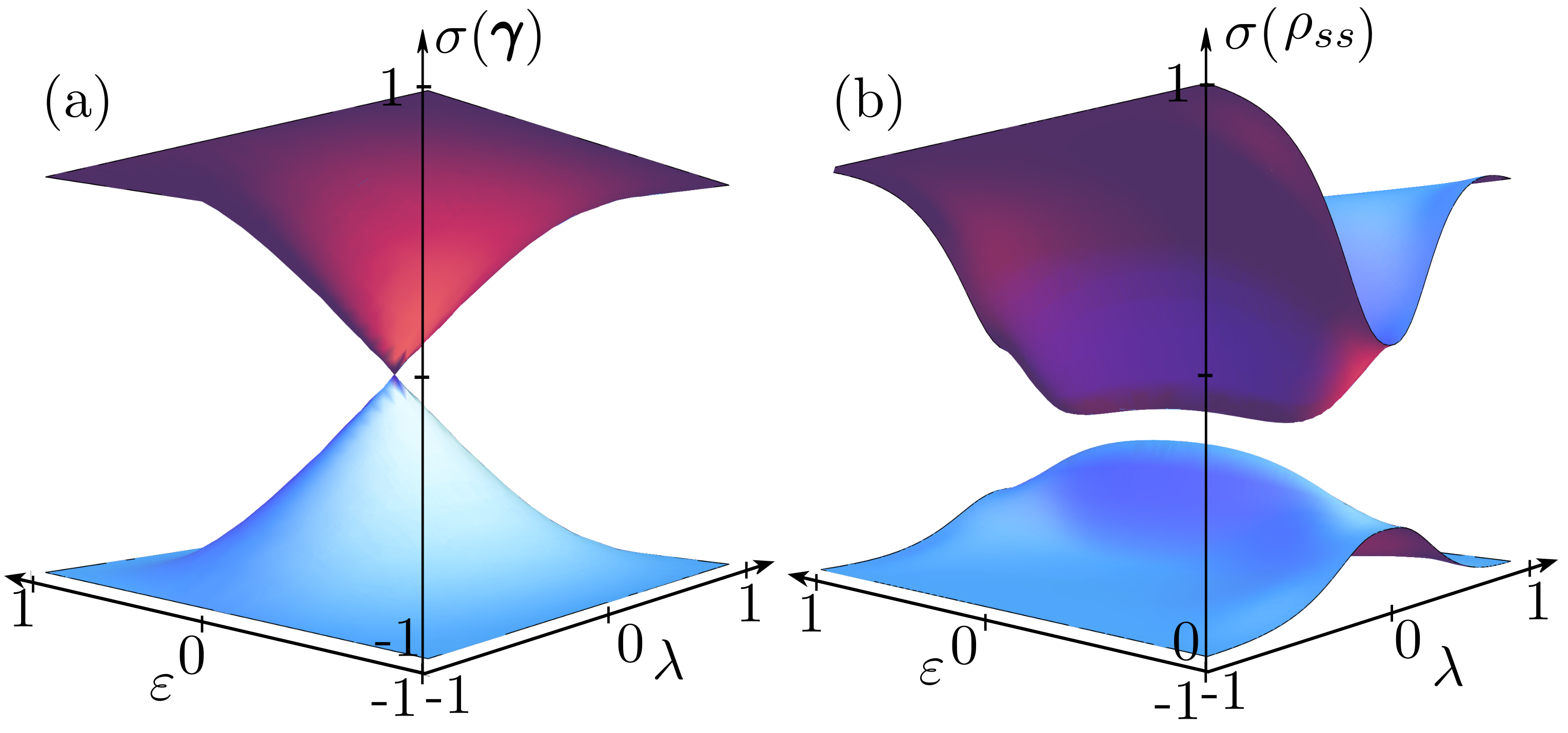, width=0.48\textwidth}
\caption{(Color online) (a) purity spectrum of fermionic model. (b) 
Dominant values of the spectrum $\sigma(\rho_\mathrm{ss})$ of the steady-state density matrix of the spin model for additional homogeneous dephasing $\gamma_\perp =0.5\Gamma$ at $\varepsilon=0$ obtained from exact 
diagonalization of a four-site problem.}
\label{fig:gaps}
\end{figure}
%%%%%%%%%%%%%%%%%%%%%%%%%%%%%%%%%%%%%%%%%%%%%%%%%%%%%

For open gaussian systems it has been argued in \cite{Bardyn-NJP-2013} that a topological singularity, at which a topological phase transitions takes place, is connected to the closing of at least one of the two, the damping gap or the so-called purity gap. The damping gap is determined by the real part of the eigenvalues of ${\cal L}$, ${\cal L} \rho_\epsilon
= \epsilon \rho_\epsilon$, which characterize the relaxation rate of the system. The steady state is a right eigenstate of ${\cal L}$ with eigenvalue $\epsilon=0$
and the damping gap $\Delta_d$ is the distance to the next larger value of Re$[\epsilon]$. The purity gap of fermionic systems is determined by the eigenvalue spectrum of the positive semidefinite matrix $(i\boldsymbol{\gamma})^2$, where $\boldsymbol{\gamma}$ is the antisymmetric and real
steady state single-particle correlation matrix
% in Majorana representation 
\cite{Bardyn-NJP-2013}.
%In Gaussian systems all properties of the steady state are encoded in the single particle spectrum.
% Pure Gaussian states have a flat purity spectrum with all eigenvalues being equal to $1$. Mixed states have eigenvalues less than $1$
% and zero eigenvalues indicated the existence of a completely mixed subspace. 

The damping gap is difficult to obtain from TEBD simulations and we have performed exact simulations for a four-site problem for the fermionic
and non-linear spin chains.
% This is shown in  Fig.\ref{fig:singularity}(a). 
While even for such small system sizes we can already observe a quantized winding of the polarization, 
% we notice at the same time that 
the damping gap is sizable everywhere (see supplementary material for details).

For the fermion model, which is gaussian, the purity gap is shown in Fig.\ref{fig:gaps} a). One recognizes a closing of the purity gap at the
origin $(\lambda=\varepsilon=0)$  consistent with the behavior of the polarization. This suggests that our classification is in agreement with previous schemes for gaussian systems such as in \cite{Bardyn-NJP-2013}.
Since the spin system cannot be mapped to a free fermionic model the concept of the purity gap 
% as an indicator of a topological singularity 
cannot be applied here. However for small systems with four sites we can evaluate the eigenvalue spectrum of the full density matrix. We find that the two dominant eigenvalues of $\rho_{\rm ss}$ never become equal (see Fig.\ref{fig:gaps} b)).
Although these calculations can only be done for finite systems this indicates that there is no closing of a "generalized" purity gap since 
an indicator
% the closing of the purity gap as a feature 
of a topological transition is expected to
be independent of system size. 
%We do find however that at the central point $\lambda =\varepsilon=0$ several eigenvalues of $\rho_{\rm ss}$ become zero. \textbf{(MF: check!DL: This is NOT the case!)}.
Thus the conditions for a topological singularity and its nature in a dissipative system remain open questions devoted to further studies.

%%%%%%%%%%%%%%%%%%%%%%%%%%%%%%%%%%
\paragraph{Summary.—}
%%%%%%%%%%%%%%%%%%%%%%%%%%%%%%%%%%%
%
To summarize, we have generalized the notion of SPT to
steady states of one-dimensional interacting open systems. 
While topological invariants based on geometric phases such as the Zak phase known from closed systems are 
no longer applicable here, the quantized bulk transport of the Thouless pump remains a suitable indicator of topology in more general one-dimensional
open systems and the winding of the polarization defines a topological invariant. 
The polarization is a measurable quantity even in an open system
and specific detection protocols will be discussed in detail elsewhere \cite{Linzner-prep}. 

As specific examples we investigated an open fermion and spin chain with reservoir couplings that lead to particle-hole symmetric steady states. 
We showed that the symmetries of the models enforce a non-trivial quantized 
winding of the many-body polarization upon cyclic parameter variations
which characterizes a topological Thouless pump. 
In the presence of inversion symmetry distinct phases with quantized values of the many-body polarization emerge which defines symmetry protected topological order.
We have shown that the polarization winding is robust to Hamiltonian disorder as well as to additional dephasing or moderate particle losses. 

The 
% relation between the existence of topological order and the underlying symmetries of the system as well as 
conditions for a topological singularity remain an
open question which requires further investigation. An interesting future avenue is the extension to higher spatial dimensions which may provide
a way to realize topologically protected transport in an open system that is robust against particle losses.

\paragraph{acknowledgment.—}
We acknowledge useful discussions with E. Demler, S. Diehl, S. Huber, M. Koster and E. v. Nieuwenburg. FG acknowledges financial support
from the Gordon and Betty Moore foundation.

\section{Supplementary}

\paragraph{Numerical method.—}
Aside from the special path, as seen in Fig.1 b) (main text), efficient numerical methods are needed to get insight into the dynamics of large systems. For this purpose we utilize an extension of the time-evolving-block-decimation algorithm (TEBD)\cite{Vidal2008,Vidal2004}. This method was originally conceived for efficient unitary time evolution of pure states in closed system and relies on the controlled truncation of the underlying Hilbert space, which is feasible for all moderately entangled states. Using small modifications however, the TEBD-algorithm can be easily extended to time evolution of mixed states in open systems.
For this we embed the mixed state as a pure state in a larger Hilbert space, the so-called Liouville space using the isomorphism
\begin{align*}
\rho=\sum_{\alpha,\beta}\rho_{\alpha,\beta}|\alpha\rangle\langle\beta|\leftrightarrow\sum_{\alpha,\beta}\rho_{\alpha,\beta}|\alpha\rangle|\beta\rangle\equiv|\rho\rangle,
\end{align*}
The state now evolves under the Schrödinger-like equation
\begin{align*}
\partial_t|\rho\rangle=\underline{\underline{\mathcal{L}}}|\rho\rangle,
\end{align*}
with $\underline{\underline{\mathcal{L}}}$ being the representation of $\mathcal{L}$ in Liouville space.

Our results are obtained in large systems (L=16) with open boundary conditions. To minimize finite size effects Eq.\eqref{eq:Resta} is evaluated within a region inside the bulk of the system (L=14) resulting in a small error because per definition periodic boundaries are required and correlations between the edges are neglected. However this error is suppressed with increasing system size and vanishes in the thermodynamic limit as in our system spin-spin correlations decay exponentially on a length scale of one lattice site.

\paragraph{Necessity of non-linear modification.—}
On the whole $\lambda=0$-axis Lindblad generators \eqref{eq:Lindblad_A} and \eqref{eq:Lindblad_B} become hermitian. Thus $\left[L_\mu,L_\mu^\dagger\right]=0$ holds and the completely mixed state
\begin{align*}
	\rho_\mathrm{M}=\bigotimes_j\frac{1}{2}\left(\ket{\uparrow}\bra{\uparrow}+\ket{\downarrow}\bra{\downarrow}\right)
\end{align*}
 is a steady state of the system.  In fact one can easily show that for this state Eq.\eqref{eq:Resta} is not defined as the argument of the logarithm becomes zero. 
 %As the behavior of the polarization becomes non-analytic one could conclude that a completely mixed steady state is a necessary condition for the existence of a topological singularity.
%To test this conjecture we 
To prevent a completely mixed steady state on the $\lambda=0$ axis we add non-linear terms \eqref{eq:nonlin_A} and \eqref{eq:nonlin_B} to the original model. 
% However a quantized winding of the polarization is still observed under this modification.
%As the polarization still shows a winding, we conclude that although the completely mixed state results in singular behavior of the polarization, it is not a necessary condition for the existence of a topological singularity.
%\item{discussion of degeneracy and necessity on nl terms}
%\item{details on damping gap and purity including old fig.4}

\begin{figure}[H]
\centering
\epsfig{file=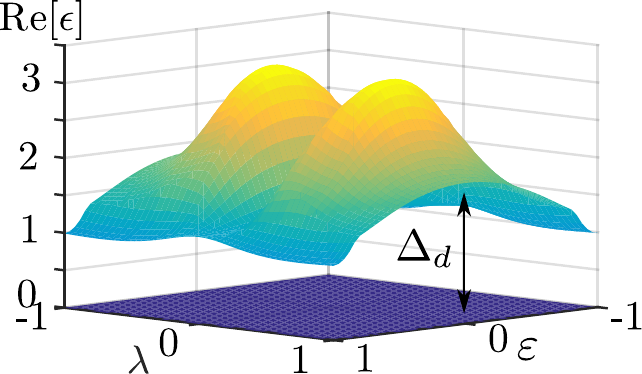, width=0.4\textwidth}%\epsfig{file=purity.pdf, width=0.3\textwidth}
\caption{(Color online) 
Two lowest eigenvalues of $\mathcal{L}$ as function of reservoir parameters $\lambda$ and $\varepsilon$ with dephasing $\gamma_\perp=0.5\Gamma$. One finds no closure of the damping gap.  }
\label{fig:damping}
\end{figure}

%
\paragraph{Damping gap.—}
Although not attainable by TEBD, we can calculate the exact damping spectrum of a small four-site spin chain, as can be seen in Fig.\ref{fig:damping}. In order to consider a most general case, while a quantized winding of the polarization is still observable, we include an additional dephasing. As can be seen from Fig.\ref{fig:damping}  there is always a finite damping gap.

$\\$

\end{document}